\documentclass[aps,prl,preprint,groupedaddress]{revtex4-1}

\usepackage{graphicx}
\usepackage{amsmath}
\usepackage{color}
\usepackage{hyperref}

\begin{document}


\title{Interaction and decay of Kelvin waves\\
 in the Gross-Pitaevskii model}



\author{Davide Proment}
\email{davideproment@gmail.com}
\homepage{http://www.uea.ac.uk/~xne12yku/}
\affiliation{School of Mathematics, University of East Anglia, Norwich Research Park, Norwich, NR4 7TJ, United Kingdom}


\author{Carlo F. Barenghi}
\affiliation{School of Mathematics and Statistics and Joint Quantum Centre (JQC) Durham-Newcastle, Newcastle University, Newcastle upon Tyne NE1 7RU, UK}

\author{Miguel Onorato}
\affiliation{Dipartimento di Fisica, Universit\`{a} degli Studi di Torino, Via Pietro Giuria 1, 10125 Torino, Italy}
\affiliation{INFN, Sezione di Torino, Via Pietro Giuria 1, 10125 Torino, Italy}



\date{\today}

\begin{abstract}
By solving numerically the governing Gross-Pitaevskii equation, we study
the dynamics of Kelvin waves on a superfluid vortex.
After determining the dispersion relation, we monitor the
turbulent decay of Kelvin waves with energy initially concentrated
at large length scales. At intermediate length scales, we find that
the decay is consistent with scaling predictions of theoretical models.
Finally we report the unexpected presence of large-length scale phonons
in the system. 
\end{abstract}


\maketitle


%
%
{\it Introduction.}
Hydrodynamic turbulence may occur in any fluid,
and is characterised by energy transfer throughout the length scales.
From a dynamical point of view, it is an out-of-equilibrium state
of a large number of degrees of freedom; its complete description is
possible only using a statistical approach
\cite{frisch1995t, zakharov41kst}.
The main properties of turbulence are
understood in terms of Richardson's energy cascade.
Energy, initially injected or stocked at large length scale $D$,
undergoes a cascade process, and it is transferred
locally through the scales in the so-called transparency window.
The cascade, driven by the nonlinear term of the governing Navier-Stokes
equation, halts at a characteristic small length scale $\eta$, 
called the Kolmogorov scale. At this small scale the diffusive 
term of the equation becomes more important than the nonlinear term, and
energy is dissipated by viscous effects.

If a constant flux of energy is injected into the system and homogeneous
isotropic turbulence develops, the statistical steady-state
which is reached
is characterized by the famous Kolmogorov energy spectrum
$E(k)\sim k^{-5/3}$ (where $k$ is the wavenumber modulus) 
in the transparency window $1/D \ll k \ll 1/\eta$,
as confirmed experimentally and numerically in the past few decades
\cite{frisch1995t}.

In this respect, pure superfluid turbulence, also called quantum turbulence, 
is less understood.  The superfluid is different from an ordinary
(classical) fluid in two respects: the viscosity is zero and 
vorticity is concentrated in discrete filaments, each carrying one 
quantum of circulation $\kappa$ \cite{donnelly1991}.
Quantum turbulence is an apparently disordered tangle of such filaments.
The superfluid behaves similarly to a classical fluid,
but only at length scales larger than the mean inter-vortex 
distance $ \ell $. At such large scales,
metastable bundles of quantum vortices \cite{PhysRevB.86.104501, PhysRevLett.109.205304}
appear spontaneously in the vortex tangle and evolve like classical eddies
\cite{PhysRevLett.101.215302}.
This qualitative picture has been confirmed by the observation of the 
Kolmogorov spectrum in laboratory experiments 
\cite{0295-5075-43-1-029, salort:125102, 2013arXiv1306.6248B}
and in numerical simulations using different theoretical models
\cite{PhysRevLett.89.145301, nore1997dkt, PhysRevLett.94.065302, 2013arXiv1306.6248B}.
However, how energy is transferred to smaller scales and finally dissipated
by phonon emission
\cite{PhysRevB.64.134520}
is still a matter of debate. 

It is widely agreed that the
major contribute to this energy transfer to smaller scales arises from
the interaction between helical vortex oscillation modes, called
Kelvin waves
\cite{thomson1880, donnelly1991, 2012arXiv1210.5194F}. 
It is thought that a Kelvin wave cascade process, similar
to the classical Richardson cascade, takes place, in which the final
energy sink is acoustic rather than viscous.
In the last decade competing theories have been proposed to explain
how Kelvin waves interact, and to account for the Kelvin waves cascade
different exponents have been predicted for the Kelvin
wave-action spectrum $ n(k)\sim k^{-\alpha} $.
By using a scale-to-scale balance argument, Vinen {\it et al.} 
\cite{PhysRevLett.91.135301} obtained $\alpha=3$.
In the framework of wave turbulence theory 
\cite{zakharov41kst, nazarenko2011wave},
Kozik and Svistunov \cite{PhysRevLett.92.035301}
considered a $ 6 $-wave interaction process leading 
to $ \alpha=17/5=3.4$,
while L'vov and Nazarenko  \cite{springerlink:10.1134}
 considered a $4$-wave process interacting with a 
large amplitude Kelvin wave field giving $ \alpha=11/3=3.\bar{6} $.
Recently, using a tilting symmetry argument, Sonin proposed a variable 
exponent $ 3 \le \alpha \le 17/5 $  that depends on the number of
interacting Kelvin waves considered
\cite{PhysRevB.85.104516}.
Although the numerical values of the predicted exponent are
relatively close to each other (hence diffucult to distinguish
numerically \cite{PhysRevB.83.134509}), 
the prefactors predicted by these theories 
differ by orders of magnitude. Clearly, better understanding of 
Kelvin waves dynamics is crucial
to explain the large-wavenumber regime of the energy cascade in a 
pure superfluid at low temperatures.

The aim of this report is to present direct observation 
of the interaction of Kelvin waves using the Gross-Pitaevskii equation (GPE).
The GPE describes quantitatevely the dynamics of a Bose-Einstein 
condensate, and, qualitatively, models a pure superfluid.
Previous results on this problem have been obtained 
using the vortex filament model based on the 
Biot-Savart law
\cite{PhysRevB.83.134509, 2011arXiv1104.4926H}.
When using the GPE, quantities indirectly related to Kelvin waves
have been monitored, such as the incompressible kinetic energy spectrum
\cite{yepez:084501, PhysRevB.84.054525, PhysRevE.83.066311, Proment:2012rt}.
The GPE has three advantages on the vortex filament model.
Firstly, it resolves scales that are comparable to the vortex core;
secondly, it includes the generation of sound waves (phonons); finally,
it includes vortex reconnection (hence there is no ambiguity when 
vortex strands become close to each other). The disadvantage of the
GPE is the limited number of length scales which are available to
the numerical solution in three dimensions.
Our work on Kelvin waves is complementary to Krstulovic's one
\cite{PhysRevE.86.055301}, who 
reported evidence in favour of the L'vov-Nazarenko spectrum of the 
Kelvin waves cascade. 

{\it The model and the numerical integration.}
The GPE describes a weakly interacting Bose-Einstein condensate.
It is also used to model a generic superfluid, and,
using the Madelung transformation, can be turned
into an equation for an inviscid barotropic compressible fluid
\cite{nore1997dkt}. 
Here we consider the dimensionless GPE
\begin{equation}
i \partial_t \psi + \frac{1}{2} \nabla^2 \psi - \frac{1}{2} |\psi|^2\psi=0,
\label{eq:GPE}
\end{equation}
where the mean density is $ \langle \rho \rangle =|\psi|^2 =1$.
In these units, the healing length (the characateristic length scale) is
$\xi=1$, the sound speed is $ c=1 $, and the quantum of circulation is 
$\kappa=2\pi$.

To study isolated Kelvin wave dynamics on a single vortex line aligned
along the x-axis, we consider a computational box with periodic 
boundary conditions in x and anti-periodic (reflective) in y and z.
As we are interested in small amplitude Kelvin waves, our
discretization is $\Delta x=\xi$, $\Delta y=\Delta z=\xi/4 $; our typical
grid has resolution  $ 256\times129\times129 $ points.
This choice is a compromise between
finite size effects (due to the reflective
boundaries), the smallest scale to be resolved, and the time of the
simulations.
The GPE is integrated in time using a standard split-step method: 
the linear terms are evolved exactly in Fourier space, and the nonlinear
terms are computed exactly in physical space. The numerical error arises
from the splitting technique \cite{bao:2006zk, Proment:2012rt}.
The discrete Fourier trasform (DFT) algorithm is applied in the x-direction,
whereas in the y- and z-directions we  use the cosine 
Fourier transform (CFT), also known as real odd discrete Fourier transform 
\cite{url:fftw}.
The chosen time step $ \Delta t=5\times10^{-3} $ is  
much smaller than the fastest linear time period, which is 
$ T_{k_{\max}} \simeq 8 \times 10^{-2} $.

The technique to prepare the initial condition is the following.
The initial wave-function is defined as
\begin{equation}
\psi(x, y, z, t_0)=\Psi_{2D}\left[y-\bar{y}(x, t_0), z-\bar{z}(x, t_0)\right],
\end{equation}
where $ \Psi_{2D}(Y, Z)=\sqrt{\rho(R)} \, e^{i\theta(Y, Z)} $ is the 
two-dimensional wave-function with a quantum vortex 
positioned in the origin of the YZ-plane (perpendicular to the
x-axis), with
$ R=\sqrt{Y^2+Z^2} $,
$ \rho(R) = R^2 \left( a_1 + a_2 R^2 \right) /
(1 + b_1 R^2 + b_2 R^4) $, and
$ \theta(Y, Z)= \mathrm{atan2}(Z, Y) $. The function
$ \mathrm{atan2}(...) $ is the extension of the arctangent function 
whose principal value is in the range $ (-\pi; \pi] $; the coefficients 
$ a_1=11/32 $, $ a_2=11/384 $, $ b_1=1/3 $, and $ b_2=11/384 $ 
arise from a second order Pad{\'e} approximation \cite{berloff:2004}.
Thus, on each slice $ x={\rm const}$, a vortex is created with centre
at $ \left[ \bar{y}(x, t_0), \bar{z}(x, t_0) \right] $.
This idea of building three-dimensional vortex lines 
starting from a two-dimensional wave-function was previously used to create 
distorted vortex rings 
\cite{PhysRevA.83.045601}
and vortex knots
\cite{PhysRevE.85.036306, 2013arXiv1307.7250P}.

The Kelvin wave state of the vortex line is defined as
\begin{equation}
w(x, t_0)=\bar{y}(x, t_0)+i \bar{z}(x, t_0) = \int A(k, t_0) \, e^{i k x} dk,
\end{equation}
where  $ w(0, t_0) = w(L_x, t_0) $  to satisfy the 
periodic boundary condition in $ x $.
The complex coefficient $ A(k, t) $ is the Fourier transform 
in $ x $ of $ w(x, t) $ and the Kelvin wave-action spectrum is simply  
$ n(k, t)=|A(k, t)|^2 $.

\begin{figure}
\includegraphics[width=0.9\linewidth]{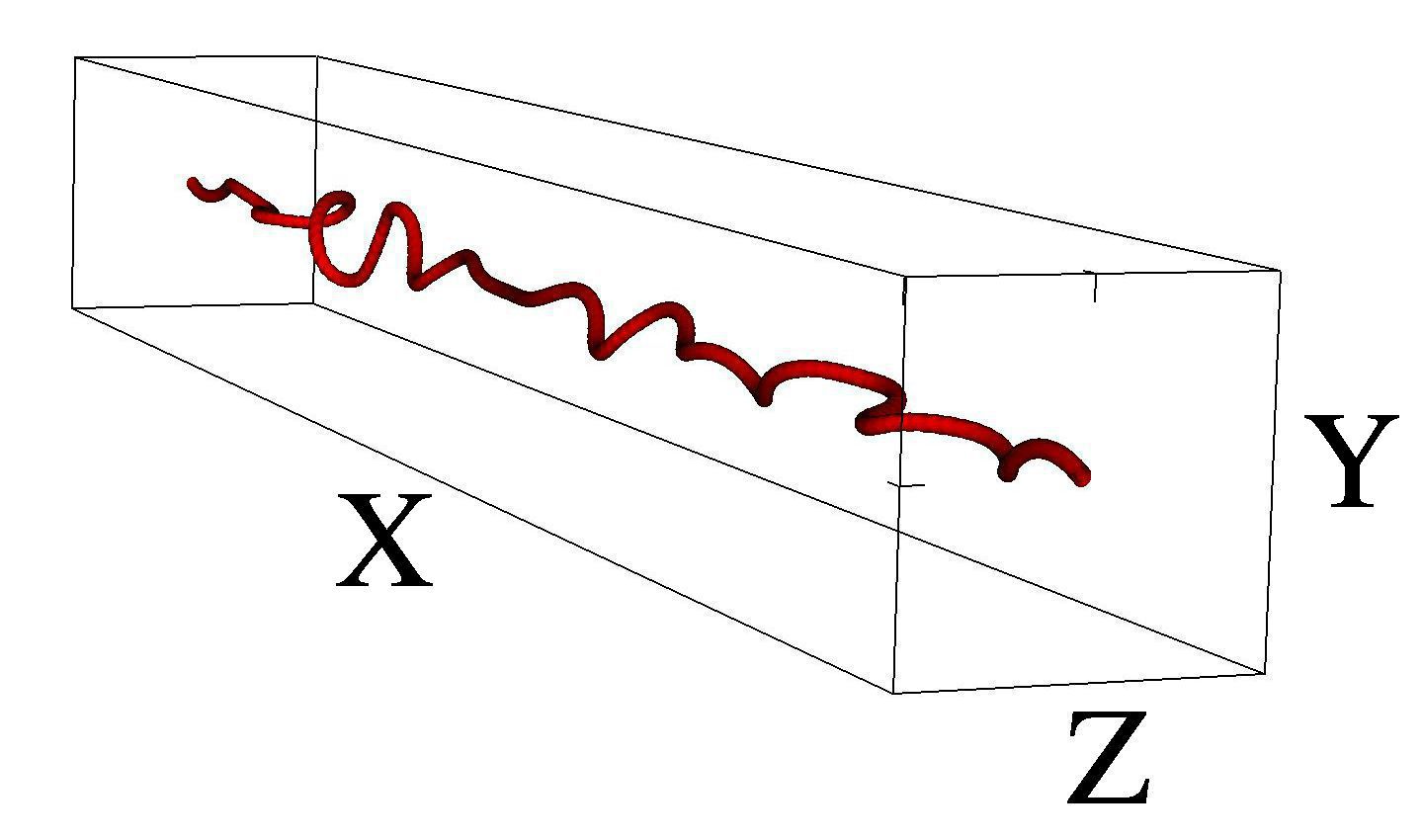}
\caption{(Colors online) Iso-surface of the density field at the threshold 
level $ \rho_{th}=0.2 $ showing the initial condition of a Kelvin wave state 
characterized
by a Gaussian spectrum peaked at large scales with random phases 
(corresponding initial spectrum shown in Fig. \ref{fig:KWSpectra}).
\label{fig:example}}
\end{figure}

All our
initial conditions are defined from a Kelvin wave spectrum,
setting $ A(k, t_0)=\sqrt{n(k, t_0)} \exp{\left[i \varphi(k)\right]} $ with
$ \varphi(k) $ randomly distributed in $ [0, 2\pi) $, and then going back to 
physical space to obtain $ w(x, t_0) $.
An example of a Kelvin wave state characterized by a Gaussian spectrum 
peaked at large scales with random phases is shown in Fig. \ref{fig:example}.

How to identify the Kelvin wave state during the time evolution?
We compute the vortex centre position on each slice $ x={\rm constant}$ 
using a simple method of weighted centre of mass below a chosen density 
threshold $ \rho=\rho_{th} $. 
The resulting coordinates of the vortex centre are
\begin{equation}
\bar{y} = \frac{\int y \, \bar{\rho} \, H\left(\bar{\rho}\right) dS}
{\int \bar{\rho} \, H\left(\bar{\rho}\right) dS}, \,\,\,
\bar{z} = \frac{\int z \, \bar{\rho} \, H\left(\bar{\rho}\right) dS}
{\int \bar{\rho} \, H\left(\bar{\rho}\right) dS},
\end{equation}
with $ \bar{\rho}=\rho_{th}-\rho $, $ H(...) $ is Heaviside's step 
function and $ dS=dxdy $.
By setting $ \rho_{th}=0.3$ and assuming axisymmetric 
density around the vortex core, we estimate the vortex centre with
accuracy an order of magnitude bigger than the size of a discretization grid cell,
$ \Delta y \Delta z $.
For completeness, we mention that a more sophisticated but numerically 
expensive technique is described in  \cite{PhysRevE.86.055301}.

{\it The dispersion relation.}
The dispersion relation is the most important quantity for Kelvin waves
dynamics. The relation describes
how each Kelvin wave component evolves independently  of the other.
The 
dispersion relation obtained
on the thin-filament approximation is \cite{donnelly1991}
\begin{equation}
\omega(k)=\frac{\kappa}{2 \pi \, a_0^2} 
\left[ 1 - \sqrt{1 + |k| a_0  \frac{K_0 \left( |k| a_0 \right)}{K_1 \left( |k| a_0 \right)}} \right] \, ,
\label{eq:dispersion}
\end{equation}
where $ a_0 $ is the vortex core cut-off parameter and $ K_n(...) $ is a modified 
Bessel function of order $ n $.
In the limit of long Kelvin wavelength $ |k| a_0 \ll 1 $ it is well approximated by 
$ \omega(k) \simeq -\kappa k^2 / 4\pi \left[ \log\left( 1/ |k| a_0 \right) - c \right] $,
with $ c=0.116... $ at a first order correction.
A similar relation was derived starting from the GPE rather than the
Biot-Savart law in the same long-wavelength limit setting
$ a_0=\xi $ \cite{pitaevskii1961vortex}; the most recent estimate of
the correction in this case is given by $ c=0.003187 $  \cite{roberts2003vortex}. 

The dispersion relation is determined in the following way.
We consider an initial condition in which many Kelvin wave modes are
excited, and compute the evolution to determine one period; in this way
we determine the relation between frequency and wavenumber $k$.

We consider an initial energy equipartioned spectrum 
$ n(k, t_0)=T/\omega(k) $,
where $ T $ represents the ``temperature'' of the Kelvin waves, and,
for simplicity, the dispersion $ \omega(k) = \kappa k^2/4\pi $ does not 
include either the logarithmic or the constant correction.
We are free to choose $ T $ in order to initially satisfy the quasi-small 
steepness condition $ |k| A(k, t_0) \le 1 $ and to make sure that the
excited vortex line is far enough from the reflective boundaries of the
computational domain.
We thus set $ T=1 $ and phases randomly distributed in
$ [0, 2\pi) $ and the resulting 
initial condition has averaged steepness $ |k| A(k, t_0) \simeq 0.4 $.

\begin{figure}
\includegraphics[width=0.9\linewidth]{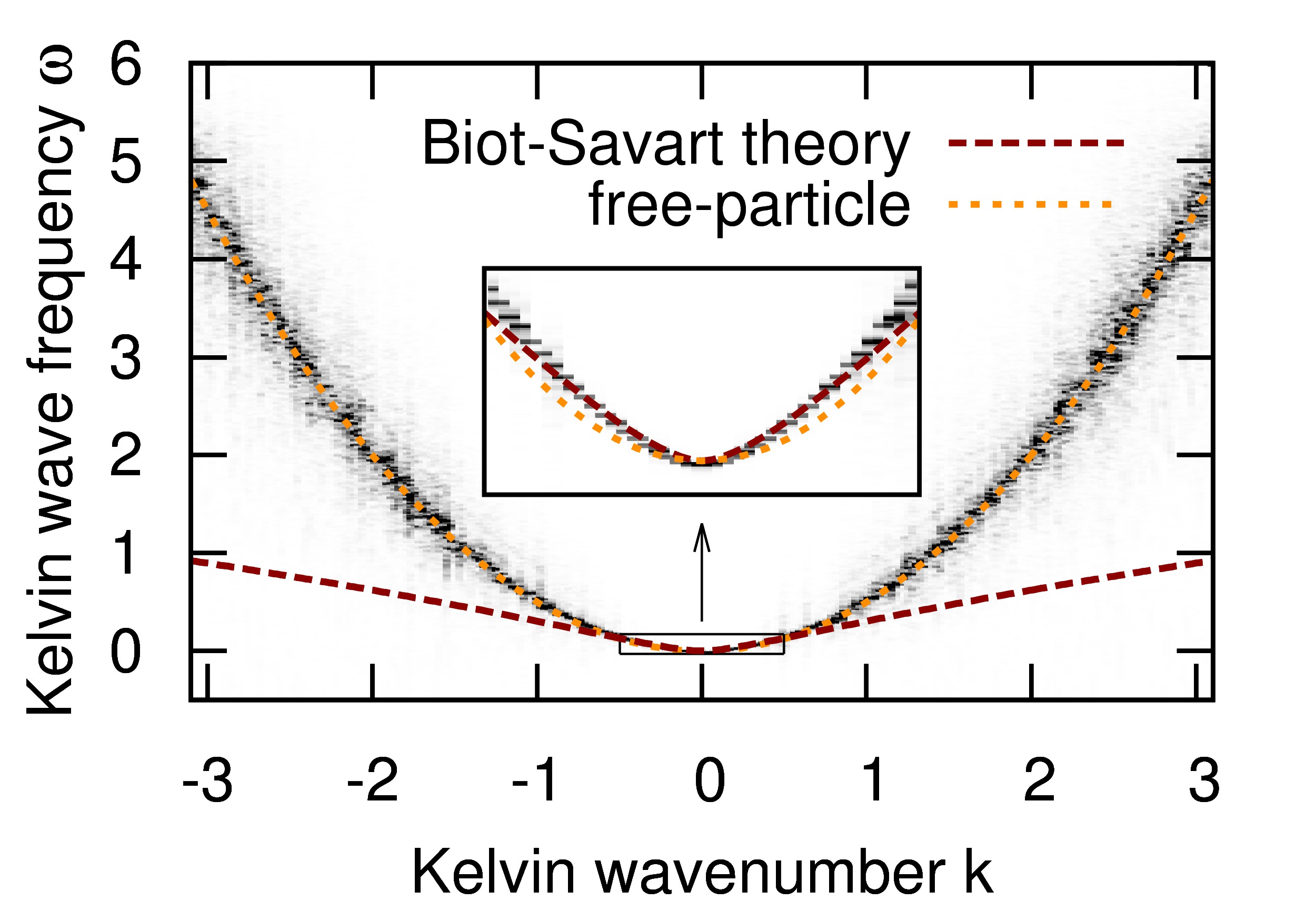}
\caption{
(Color online) Kelvin wave dispersion relation $\omega$ vs $k$
evaluated for an initial condition characterized by the Kelvin wave 
spectrum $ n(k, t_0)=T/\omega(k) $ with $ T=1 $, $ \omega(k) = \kappa k^2/4\pi $, 
and phases 
randomly distributed in $ [0, 2\pi) $.
The blue dashed line corresponds to the theoretical prediction coming from the Biot-Savart filament model (\ref{eq:dispersion}) with $ a_0=\xi $.
The red line is the estimated free-particle dispersion $ \omega_{est}(k)=\kappa k^2/ 4\pi $.
Inset: a zoom at the low Kelvin wavenumbers.
\label{fig:dispersionRelation}
}
\end{figure}

The dispersion relation which we obtain evolving the governing GPE model for 
sufficiently long turnover times is shown
in Fig. \ref{fig:dispersionRelation}.
Equation (\ref{eq:dispersion}), widely used in the literature,
fits well the computed dispersion only for 
small wavenumbers, while for $ |k| \xi \ge1 $ the simpler expression
$ \omega_{est}(k)=\kappa k^2/4\pi=k^2/2 $ showing free-particle behaviour 
better fits the data.
The failure of dispersion (\ref{eq:dispersion}) is expected at the scale $ k=1/a_0 $ as
Biot-Savart assumes that the vortex core size $ a_0 $ is infinitesimal compare to 
the Kelvin wave amplitudes and Fig. \ref{fig:dispersionRelation} shows exactly the 
crossover between the two regimes.
Our result is consistent with the numerical results of Roberts
\cite{roberts2003vortex},
and does not exhibit the negative frequency shift $ \omega_0 $ observed in
\cite{PhysRevLett.101.020402, PhysRevA.79.033619}.
It is important to notice that the energy equipartitioned initial spectrum
is not the statistical steady state of the system as during the evolution 
we observed changes in its shape.   

\begin{figure}
\includegraphics[width=0.9\linewidth]{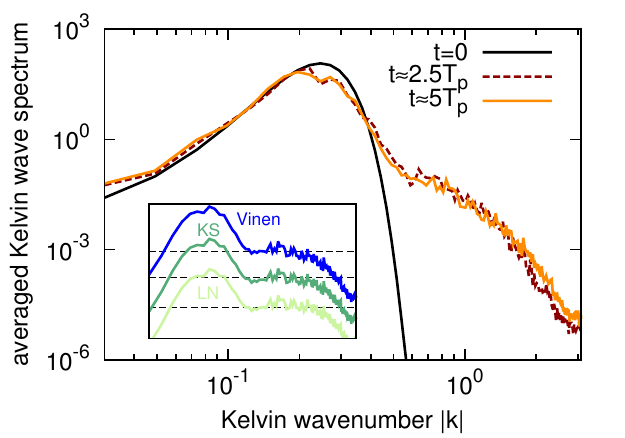}
\caption{(Color online) Averaged Kelvin wave spectrum
$ \langle n(k, t) \rangle $ plotted vs wavenumber $k$ on a log-log scale
at different times $ t $ during the evolution. 
The spectrum evolves from the initial Gaussian
shape and approximately acquires power-law dependence
at intermediate wavenumbers ($ 0.5 \le |k| \le 2$).
Inset: the final spectrum $ \langle n(k, 5 \, T_p) \rangle $ vs wavenumber,
compensated by Vinen's ($k^{-3}$), 
Kozik and Svistunov's ($k^{-17/5}$, KS), and L'vov and Nazarenko's 
($k^{-11/3}$, LN) scalings respectively.
\label{fig:KWSpectra}}
\end{figure}

{\it The Kelvin wave decay turbulence.}
In order to understand the interaction of Kelvin waves which eventually leads
to a turbulent energy cascade, we simulate the free decay and monitor the
expected transfer of energy from large to small scales.
We set up an initial Gaussian-shaped spectrum 
$ n(k, t_0)=A/\left(\sigma\sqrt{2\pi}\right) 
\exp{\left[-(|k|-\mu)^2/(2\sigma^2)\right] } $, 
with $ A=2 $, $ \mu=10 \, \Delta k_x $, and 
$ \sigma=3 \, \Delta k_x  $, where $ \Delta k_x=2\pi/256 \, \Delta x $ being
the smallest Kelvin wave mode. With
this choice, almost all the initial energy is contained in the largest
length scales.
We perform five numerical simulations with different initial random phases;
the resulting directional- and ensemble-averaged spectral evolution
$ \langle n(k, t) \rangle $ is shown in Fig. \ref{fig:KWSpectra}.
The value of time cited in the figure refers not to the dimensionless
units used to solve the GPE, but rather to Kelvin wave turnover period 
$ T_p=2\pi/\omega_{est}(k_p) $, where 
$ k_p=\mu $ is the wavenumber at which the initial large-scale  spectrum
peaks. In these units, the nonlinear evolution is rapid: starting from the
initial concentration at the mesocales,  energy is visibly shifted to the 
largest wavenumber already at $ t\simeq 2.5 \, T_p $. During the successive
stage of the evolution, the power-law scaling appears approximately in the
wavenumber range interval $ 0.5 \le |k| \le 2$.
The inset shows the final averaged spectrum 
$ \langle n(k, 5 \, T_p) \rangle $ compensated by the scalings
predicred by Vinen {\it et. al}, Kozik and Svistunov, and L'vov and Nazarenko.
It is apparent that, as found using the vortex filament model
\cite{PhysRevB.83.134509}, all the predicted scalings are close to the
numerical results.

\begin{figure}
\includegraphics[width=0.9\linewidth]{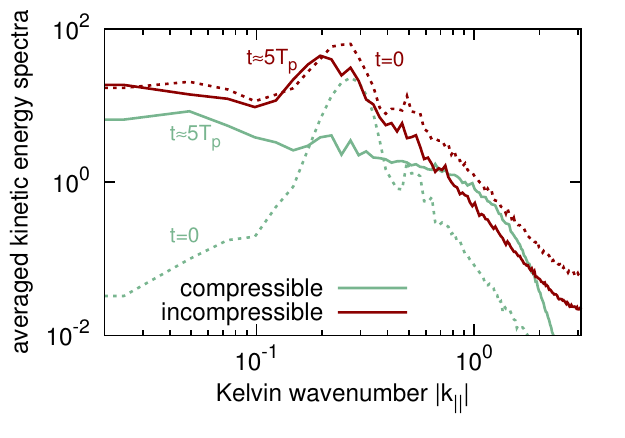}
\caption{(Color online) Averaged parallel compressible and incompressible 
kinetic energy spectra $ \langle E^c(k_{\parallel}, t) \rangle $ and 
$ \langle E^i(k_{\parallel}, t) \rangle $ vs wavenumber $k$, plotted on
log-log scale at different times.
\label{fig:kineticSpectraH}}
\end{figure}

\begin{figure}
\includegraphics[width=0.9\linewidth]{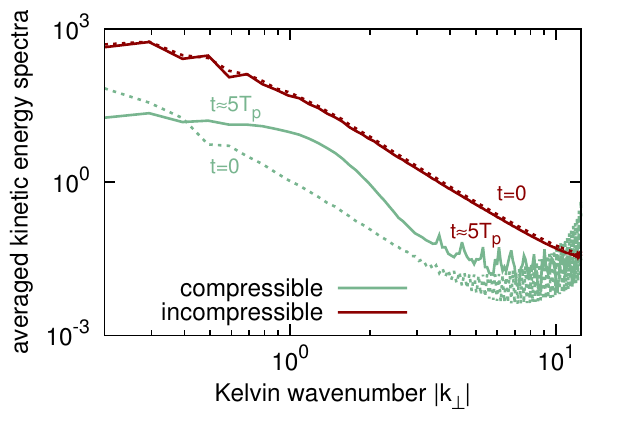}
\caption{(Color online) Averaged perpendicular compressible and incompressible 
kinetic energy spectra $ \langle E^c(k_{\perp}, t) \rangle $ 
and $ \langle E^i(k_{\perp}, t) \rangle $ vs wavenumber $k$, plotted
on log-log scale at different times.
\label{fig:kineticSpectraR}}
\end{figure}

To gain physical insight into the Kelvin waves interaction, we compute the
spectrum of the incompressible kinetic energy, which is only a fraction 
of the total (conserved) energy of the system. 
Following \cite{nore1997dkt},
we split the superfluid's kinetic energy  into compressible and 
incompressible parts,
associated to the presence phonons and quantum vortices respectively. 
Fig.s \ref{fig:kineticSpectraH} and \ref{fig:kineticSpectraR} display
the (directional- and ensembe-averaged) one-dimensional 
kinetic energy spectra, where we distinguish between
the directions parallel ($ k_{\parallel} $) and perpendicular 
($ k_{\perp} $) to the unperturbed vortex line (in practice,
$ k_{\parallel}=k_x $, whereas $ k_{\perp} $ is the 
Fourier transform of the radius coordinate in the $YZ$ plane, 
averaged over the angular coordinate).


The numerical results suggest the following three conclusions.
Firstly, it is clear in Fig. \ref{fig:kineticSpectraH} that 
the parallel compressible energy component (initially negligible 
at all length scales compared to the incompressible component),
at the final time $t=5 \, T_p$ becomes dominant in the range $ |k| \xi \ge 0.7 $.
We conclude that energy is transferred from Kelvin waves to phonons
mostly at these smaller scales.

Secondly, the perpendicular incompressible energy component,
shown in Fig. \ref{fig:kineticSpectraR},
does not evolves in time in a significant way. Likely, this quantity
is completely dominated by the vortex core and
not by the different Kelvin wave modes which are excited long the
vortex line.

Thirdly, both parallel and perpendicular compressible energy components 
seem to equilibrate to a power-law behavior
at scales $ |k| \xi \le 1 $, but are strongly reduced at smaller scales.
This result is somehow unexpected, because the scenario which is described in
the literature seems to imply that phonons should be mainly produced
at length scales which are much smaller than the length scales which
initially contained the energy.
We do not know if the mechanism which generates large scale phonons 
is due to the presence of the oscillating vortex or to phonon interaction
\cite{PhysRevA.80.051603}.
We have verified that the effect is present also when we start
from a different initial condition (smaller amplitude and
less compressible energy).

{\it Conclusions.}
We have performed numerical simulations of interacting Kelvin waves
using the GPE and an idealized vortex configuration which consists
of only one perturbed vortex line. The method which we have developed
to numerically produce the desired Kelvin wave spectrum on the 
vortex line and track the vortex core position is general and can be used
for other studies.

Our first step was to determine the dispersion relation of Kelvin
waves, starting from an initial condition where all Kelvin modes are 
excited and (in the first approximation) energy is equipartioned.
As expected, we have found that the dispersion relation often used
in the literature (Eq.~\ref{eq:dispersion}) is valid only for long
waves ($ |k| a_0 < 0.2$, where we set $ a_0=\xi $).

The second step was to determine the turbulent decay of an initial
spectrum with most of the energy stored at large scales.
By running different simulations with different initial random phases
and ensemble-averaging,
we have obtained a power-law spectrum which is consistent with
the predictions of the competing theories for the Kelvin wave turbulence.
Unfortunately, as the inertial range is too narrow, we are unable to 
draw any strong conclusions about which theory is more consistent with
our data (unlike what done in \cite{PhysRevE.86.055301}).

The thrid step was to extract the parallel and perpendicular
compressible and incompressible contributions to the
kinetic energy.  We found that, at small length scales,
the compressible parallel component 
becomes stronger than the incompressible componnent,
confirming that phonons play a crucial role in the decay.
Unexpectedly, we observe the formation of large scale phonons 
in both energy components.


%
%
%

%
%
{\it Acknowledgments.}
The authors are grateful to
A.W. Baggaley, G. Boffetta, R. H{\"a}nninen, G. Krstulovich, 
J. Laurie, F. De Lillo, 
S. Nazarenko, and H. Salman for discussions and suggestions.

\bibliography{references}

\end{document}